\documentclass{emulateapj} %% \\for the arxiv
\usepackage{epsfig}
\bibdata{grbPaper}
\begin{document}

\title{The First Limits on the Ultra-High Energy Neutrino Fluence from Gamma-Ray Bursts}

\author{A.~G.~Vieregg\altaffilmark{1,10}}
\author{K.~Palladino\altaffilmark{2}}
\author{P.~Allison\altaffilmark{2}}
\author{B.~M.~Baughman\altaffilmark{2}}
\author{J.~J.~Beatty\altaffilmark{2}}
\author{K.~Belov\altaffilmark{1}} 
\author{D.~Z.~Besson\altaffilmark{3}}
\author{S.~Bevan\altaffilmark{4}}
\author{W.~R.~Binns\altaffilmark{5}}
\author{C.~Chen\altaffilmark{6}}
\author{P.~Chen\altaffilmark{6}}
\author{J.~M.~Clem\altaffilmark{7}}
\author{A.~Connolly\altaffilmark{2}}
\author{M.~Detrixhe\altaffilmark{3}}
\author{D.~De~Marco\altaffilmark{7}}
\author{P.~F.~Dowkontt\altaffilmark{5}}
\author{M.~DuVernois\altaffilmark{8}}
%B.~Fox$^1$}
\author{P.~W.~Gorham\altaffilmark{8}}
\author{E.~W.~Grashorn\altaffilmark{2}}
%N.~Griffith$^2$}
%M.~Hannawald$^1$}
\author{B.~Hill\altaffilmark{8}}
\author{S.~Hoover\altaffilmark{1}}
\author{M.~Huang\altaffilmark{6}}
\author{M.~H.~Israel\altaffilmark{5}}
\author{A.~Javaid\altaffilmark{7}}
%J.~G.~Learned$^1$}
\author{K.~M.~Liewer\altaffilmark{9}}
\author{S.~Matsuno\altaffilmark{8}}
\author{B.~C.~Mercurio\altaffilmark{2}}
\author{C.~Miki\altaffilmark{8}}
\author{M.~Mottram\altaffilmark{4}}
\author{J.~Nam\altaffilmark{6}}
\author{R.~J.~Nichol\altaffilmark{4}}
\author{A.~Romero-Wolf\altaffilmark{8}}
\author{L.~Ruckman\altaffilmark{8}}
\author{D.~Saltzberg\altaffilmark{1}}
\author{D.~Seckel\altaffilmark{7}}
\author{G.~S.~Varner\altaffilmark{8}}
\author{Y.~Wang\altaffilmark{6}}

\altaffiltext{1}{Dept. of Physics and Astronomy, University of California, Los Angeles, CA 90095}
\altaffiltext{2}{Dept. of Physics, Ohio State University, Columbus, OH 43210.}
\altaffiltext{3}{Dept. of Physics and Astronomy, University of Kansas, Lawrence, KS 66045.}
\altaffiltext{4}{Dept. of Physics and Astronomy, University College London, London, United Kingdom.}
\altaffiltext{5}{Dept. of Physics, Washington University in St. Louis, MO 63130.}
\altaffiltext{6}{Dept. of Physics, National Taiwan University, Taipei, Taiwan.}
\altaffiltext{7}{Dept. of Physics, University of Delaware, Newark, DE 19716.}
\altaffiltext{8}{Dept. of Physics and Astronomy, University of Hawaii at Manoa, Honolulu, HI 96822.}
\altaffiltext{9}{Jet Propulsion Laboratory, Pasadena, CA 91109. }
\altaffiltext{10}{Current Address: Harvard-Smithsonian Center for Astrophysics, Cambridge, MA 02138; \email{avieregg@cfa.harvard.edu}}

\begin{abstract}
We set the first limits on the ultra-high energy (UHE) neutrino fluence at energies greater than $10^{9}$~GeV from gamma-ray bursts (GRBs) 
based on data from the second flight of the ANtarctic Impulsive Transient Antenna (ANITA). 
During the 31 day flight of ANITA-II, 26 GRBs were recorded by Swift or Fermi.
Of these, we analyzed the 12 GRBs which occurred during quiet periods when the payload was 
away from anthropogenic activity.  In a blind analysis, we observe 0 events on a total background 
of 0.0044 events 
in the combined prompt window for all 12 low-background bursts.  We also observe 0 events from the 
remaining 14 bursts.  We place a
90\% confidence level limit on the $E^{-4}$ prompt neutrino fluence between
10$^{8}$~GeV$<$E$<$10$^{12}$~GeV of $E^4\Phi$=2.5$\times 10 ^{17}$~GeV$^3$/cm$^2$ from GRB090107A.  
This is the first reported limit on the UHE neutrino fluence from GRBs above 10$^{9}$~GeV, and the 
strongest limit above 10$^{8}$~GeV.
\end{abstract}


\section{Introduction}
Gamma-ray bursts (GRBs) are the most powerful explosions in the universe, and are thus considered
to be a possible source of ultra-high energy (UHE) cosmic rays.  Short-duration bursts are believed to be 
% A standard theory
%describes short-duration bursts as a 
a result of the collision of two compact objects
and long-duration bursts are thought to be beamed emission from the collapse of a high-mass star into a black hole.
See~\citet{meszaros1, meszaros2} for reviews of the basic theories of
GRBs.  In the widely-accepted fireball shock model, relativistic plasma in a jet collides either
with the surrounding material or with the outflow itself, producing the observed gamma-ray
prompt emission through synchrotron and inverse Compton scattering of electrons~\citep{meszarosRees1}. Protons are
also thought to be accelerated in these shocks via the Fermi mechanism~\citep{wick,dermer}.  
These UHE protons then interact
with the photons, going through a $\Delta^+$ resonance and producing charged pions which decay, 
yielding UHE neutrinos~\citep{becker,halzenHooper}.  
The first calculations of this prompt UHE neutrino emission use an E$^{-2}$ 
proton injection spectrum with energies up to E=$10^{11}$~GeV, and predict an E$^{-4}$ 
neutrino spectrum in the UHE regime, with the steepening of the spectrum 
due to synchrotron cooling of the UHE pions~\citep{wb1, wb2, wb3,alvarez}.

The detection of UHE neutrinos from GRBs would support their identification as the sources of the highest
energy cosmic rays, a longstanding mystery in particle astrophysics.  Previous searches for neutrino production
have been perfomed by the IceCube~\citep{icecube,icecube40} and RICE collaborations~\citep{rice},
but this is the first search for UHE neutrinos from GRBs above 10$^{9}$~GeV.

\section{The ANITA Instrument}
A full description of the
ANITA-I instrument can be found in~\citet{anitaInstrument}, and a description of instrument modifications
for ANITA-II is in~\citet{vieregg}~and~\citet{anita2}.  Briefly, 
the ANITA experiment is a NASA Long Duration Balloon experiment that searches for coherent, impulsive, broadband 
radio emission (200-1200~MHz) from electromagnetic showers induced by UHE neutrinos interacting in the 
Antarctic ice sheet~\citep{anitaInstrument, askaryan}.  The second flight of the ANITA experiment launched
on 2008 December 21, flew for 31 days, 28.5 of which were live days, and recorded
over 26~million triggers.  Forty quad-ridged, dual-polarization
horn antennas search for radio impulses which could be caused by neutrino interactions 
in the ice sheet.  The trigger requires coherent power in neighboring antennas,
and the threshold is limited by thermal-noise emission from the ice.  
Over 98.5\% of recorded events
were fluctuations of thermal noise.  The trigger is designed to optimize efficiency
on neutrino-like signals: vertically-polarized, broadband impulses.  
Signals from each polarization of each antenna 
are recorded in a 100~ns window for each triggered event, 
allowing for directional determination on an event-by-event 
basis using interferometric techniques, also described at length in~\citet{anitaInstrument}.  
ANITA is most sensitive to neutrinos which come from between the horizon and a payload 
elevation angle (angle above the horizontal) of $-25^{\circ}$.

%A recent search for neutrino events during the second flight of the ANITA experiment~\cite{anita2} found
%two neutrino candidate events on a background of 0.97$\pm$0.42, and set the strongest limit
%to date on the astrophysical UHE neutrino flux above 10$^{18}$~eV.

\section{Data Analysis}
We are able to construct a more sensitive search for UHE neutrinos from GRBs compared to the previously
reported diffuse UHE neutrino search with ANITA-II~\citep{anita2} because the short time window
given by the burst duration dramatically reduces background 
in the signal region, allowing us to lower our analysis threshold and look for 
very weak signals which also have a time and direction correlation with the observed GRB.

There are two sources of background for an ANITA neutrino search.  The first is 
thermal-noise fluctuations, which are easily removed with a set of cuts on the strength 
of waveform correlation among neighboring antennas and the signal strength.  The second
is man-made noise, which can be removed because it 
tends to cluster with locations of known human activity and with
other events.  The details of event reconstruction, thermal-noise rejection, and
man-made noise rejection are discussed in~\citet{vieregg}~and~\citet{anita2}.  Compared to the 
diffuse neutrino search, we loosen numerical values of thermal-noise cuts.
The cuts against man-made noise remain the same.
\begin{table*}[htb2!]
\caption{\label{tab:info}List of the 12 GRBs included in the blind analysis
}
\begin{center}
\begin{tabular}{lcccc}
\hline
GRB & Date \& Time (UTC)  & Right Ascension & Declination & Payload Elevation\\
&&&& Angle (degrees)\\
\hline
081228 & 2008 Dec 28 01:17:40 & $2^{\mathrm{h}}37^{\mathrm{m}}50^{\mathrm{s}}.94$ & $30^{\circ}51{\mathrm{'}}10~50{\mathrm{''}}$ & -41.1\\
081229 & 2008 Dec 29 04:29:01.88 & $11^{\mathrm{h}}22^{\mathrm{m}}0^{\mathrm{s}}$ & $55^{\circ}6{\mathrm{'}}0{\mathrm{''}}$ & -55.9\\
081230 & 2008 Dec 30 20:36:12 & $2^{\mathrm{h}}29^{\mathrm{m}}19^{\mathrm{s}}.51$ & $-25^{\circ}8{\mathrm{'}}49~95{\mathrm{''}}$ & 28.8\\
081231 & 2008 Dec 31 03:21:01.93 & $14^{\mathrm{h}}35^{\mathrm{m}}0^{\mathrm{s}}$ & $-38^{\circ}43{\mathrm{'}}0{\mathrm{''}}$ & 47.0\\
090102 & 2009 Jan 2 02:55:36 & $8^{\mathrm{h}}32^{\mathrm{m}}58^{\mathrm{s}}.54$ & $33^{\circ}6{\mathrm{'}}51~10{\mathrm{''}}$ & -26.3\\
090108B & 2009 Jan 8 07:43:23.36 & $0^{\mathrm{h}}15^{\mathrm{m}}0^{\mathrm{s}}$ & $-32^{\circ}12{\mathrm{'}}0{\mathrm{''}}$ & 42.9\\
090109 & 2009 Jan 9 07:58:29.49 & $8^{\mathrm{h}}11^{\mathrm{m}}0^{\mathrm{s}}$ & $54^{\circ}48{\mathrm{'}}0{\mathrm{''}}$ & -59.2\\
090111 & 2009 Jan 11 23:58:21 & $16^{\mathrm{h}}46^{\mathrm{m}}42^{\mathrm{s}}.14$ & $0^{\circ}4{\mathrm{'}}38~21{\mathrm{''}}$ & 1.7\\
090112A & 2009 Jan 12 07:57:23.11 & $7^{\mathrm{h}}27^{\mathrm{m}}0^{\mathrm{s}}$ & $-30^{\circ}17{\mathrm{'}}0{\mathrm{''}}$ & 23.5\\
090112B & 2009 Jan 12 17:30:15.45 & $12^{\mathrm{h}}51^{\mathrm{m}}0^{\mathrm{s}}$ & $22^{\circ}12{\mathrm{'}}0{\mathrm{''}}$ & -26.8\\
090113 & 2009 Jan 13 18:40:39 & $2^{\mathrm{h}}8^{\mathrm{m}}13^{\mathrm{s}}.63$ & $33^{\circ}25{\mathrm{'}}42~85{\mathrm{''}}$ & -25.7\\
090117B & 2009 Jan 17 08:02:02.23 & $15^{\mathrm{h}}32^{\mathrm{m}}0^{\mathrm{s}}$ & $27^{\circ}36{\mathrm{'}}0{\mathrm{''}}$ & -28.1\\
\hline
\end{tabular}
\end{center}
\end{table*}

Any prompt emission neutrino candidate events would be
vertically-polarized events which occur 
during the GRB prompt emission window 
($T_{90}$, the time over which 90\% of gamma-rays were detected), 
pass thermal-noise cuts, and are hardware-triggered
in the direction of the observed GRB~($\pm22.5^{\circ}$).  
Events are also rejected if they cluster with man-made noise 
(identified in the previous UHE neutrino search analysis~\citep{anita2})
or with locations of known human activity on the Antarctic continent.  We also searched for precursor
neutrino emission in the 100~seconds before the start of the burst~\citep{razzaque}.

We proceed in the search for GRB-coincident neutrinos using a blind method.  We set all analysis cuts
on regions of time which should contain no neutrino events, and then apply the same
cuts in the prompt and precursor emission windows.  For the purpose of setting analysis cuts, 
we chose the 55 minutes starting 1 hour before each burst and the 55 minutes starting 5 minutes
after the burst (for a total of 1 hour and 50 minutes) to be the background period for the
burst.  This allows us to use events close to the signal region in time as a background 
sample without ruling out the possibility of extended prompt or precursor neutrino emission.

\begin{figure}[tbp]
\centering
\includegraphics[width=8.0cm]{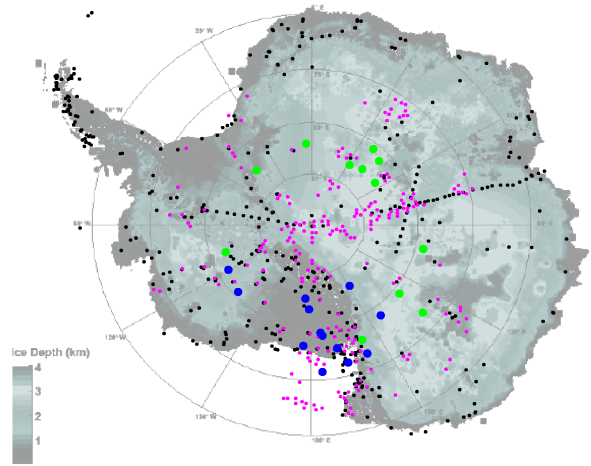}
\caption{The location of the ANITA payload during each of the 26 bursts recorded during the 
flight of ANITA-II.  Green locations are bursts with clean background periods, and blue locations are bursts with
anthropogenic noise in the background periods.  Locations of human activity are shown in black (known) 
and magenta (possible).} 
\label{fig:continent} 
\end{figure}

Of the 26 bursts observed by Swift or Fermi during the flight of ANITA-II,
only 12 bursts had background periods with a
thermal-like distribution of events.  Information from the Gamma-ray bursts Coordinates Network~(GCN,~\citet{gcn}) 
about these 12 bursts is in Table~\ref{tab:info}.  The remaining 14 bursts had significant anthropogenic 
activity in the background periods, and were removed from the analysis to reduce the risk of man-made
events occurring during the signal time window.  Figure~\ref{fig:continent} shows the location of the ANITA
payload when each of the 26~GRBs occurred.  The green circles indicate bursts with clean background
periods, while the blue circles indicate bursts with noisy background periods.  As expected, the bursts with noisy
background periods are from times when the payload is near McMurdo station and the Ross Ice Shelf, the 
part of the continent with the most human activity.

We set the analysis cuts so that if one event were found in the prompt emission signal
region for any burst, it would be a three sigma fluctuation of the expected background.  We set
the final analysis cut (on the peak value of the cross-correlation from the interferometric image, 
described in~\citet{anita2}) to allow 0.0044 background events in the total prompt signal
region for all 12 bursts.  The final analysis cuts used in this search 
are over 98\% efficient for events which trigger, even for the smallest low-SNR triggered signals, 
when tested on signal-like calibration events from a radio-frequency pulsing station at Taylor Dome.

\section{Results}
We found no events in the blind signal region for prompt emission ($T_{90}$).  
We also found no events in the precursor window (100~s before the start of each burst).  This 
is consistent with the background expectation.  We proceed to set a limit for each burst
individually on the prompt UHE
neutrino fluence using a Feldman-Cousins 90\% confidence interval, the duration of the burst,
and the acceptance calculated using an ANITA Monte Carlo simulation.  For each GRB, 
we configured the Monte Carlo to simulate a point source at the location of the burst, 
fixed ANITA at the location of the payload during the burst, and assumed an input E$^{-4}$ 
spectrum.  We investigate systematic effects on the ANITA-II diffuse neutrino limit 
due to uncertainties in neutrino cross section,
surface roughness, and birefringence in~\citet{anita2}, and
the limit presented here would be affected by approximately the same factor, $\sim10$\%.

Table~\ref{tab:fluence} contains the calculated synchrotron 
break energy (where the spectrum turns from E$^{-2}$ to E$^{-4}$), as well as predicted fluxes beyond
the break for each of the 12 GRBs used in this search.  The calculations follow
methods in~\citet{rice}, which are based on Waxman-Bahcall calculations~\citep{wb1, wb2, wb3, waxman}.
Redshift and gamma-ray flux information are available from the
GCN~\citep{gcn}.
For bursts with no redshift information, we use a redshift
of $z=2$ for flux calculations, following~\citet{icecube}.  For all bursts, 
the UHE regime is well past the break energy, leading to an $E^{-4}$ spectrum 
over the entire ANITA energy range.

There are two ways that ANITA can view the radio emission from a neutrino interacting in the ice.  The first
geometry, called a \textit{direct} observation, occurs when
ANITA observes the radio impulse directly from the interaction of an upgoing neutrino.  
The second geometry, called a \textit{reflected} observation,
occurs when ANITA sees the radio impulse reflected off of the bottom of an ice shelf (at
the sea water interface) from the interaction of a downgoing neutrino.  
Since UHE neutrinos are absorbed as they travel 
through the Earth, most of ANITA's direct events would be associated with 
neutrinos which skim across the ice.  %The optimal 
%direct observation angle, calculated using the ANITA Monte Carlo simulation, 
%is $-10^{\circ}$ elevation (measured from horizontal).  
When ANITA is over the Ross Ice Shelf, it can also 
make a reflected observation of downgoing neutrinos.

\begin{figure*}[tbp]
\centering
\includegraphics[width=19.0cm]{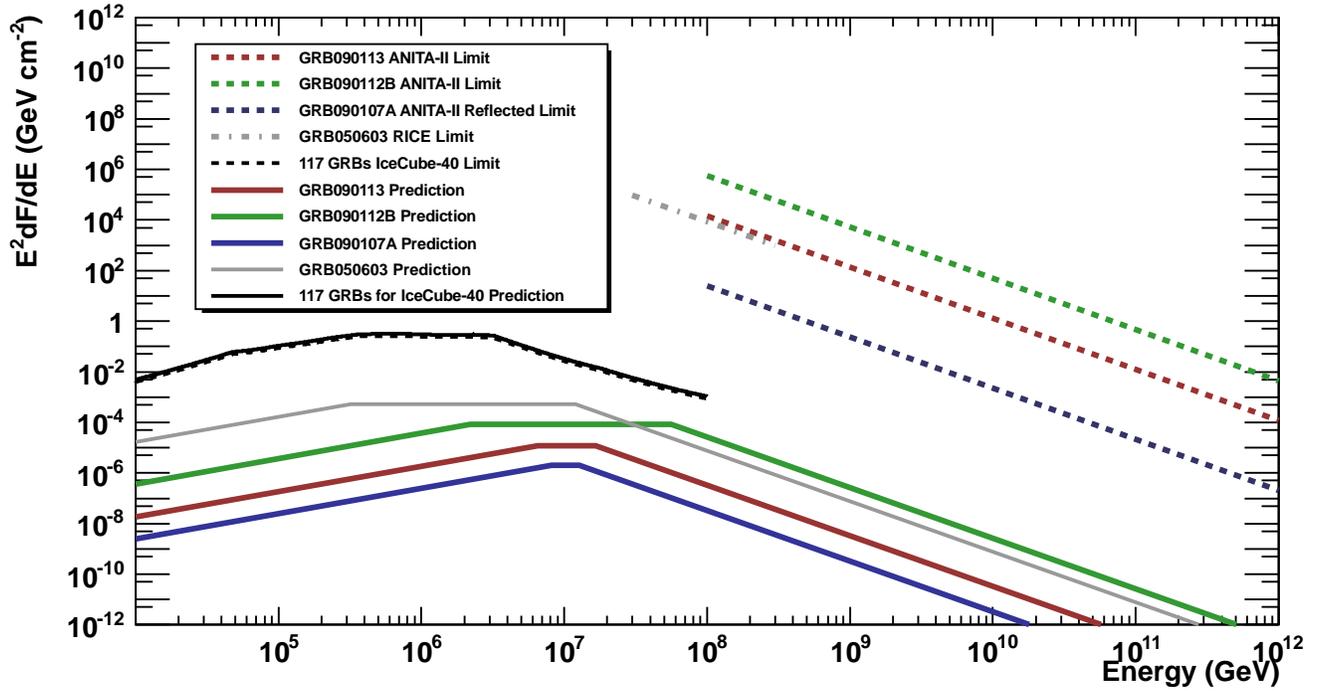}
\caption{The two best direct limits on the UHE neutrino 
fluence from the blind analysis are from GRB090113 and GRB090112B, and are 
shown with red and green dashed lines respectively.    
The best reflected limit is from GRB090107A, shown with a blue dashed line.   RICE~\citep{rice}
and IceCube~\citep{icecube40} limits are also shown.  The IceCube limit is an aggregate limit based on 117
individual GRBs, and is based on a fluence prediction from~\citet{guetta}.} 
\label{fig:limit} 
\end{figure*}

None of the 26 GRBs during the flight had a payload elevation angle
between $-25^{\circ}$ and the horizon, which is where ANITA has
the best chance of seeing direct neutrino events.  
Of the 12 GRBs used in this search, 
the most promising direct observation geometry was from GRB090113 
(an elevation angle of $-25.7^{\circ}$).  The 90\% confidence level 
fluence limit for energies 10$^{8}$~GeV$<$E$<$10$^{12}$~GeV 
is $E^4\Phi$=1.5$\times 10 ^{20}$~GeV$^3$cm$^{-2}$ from GRB090113, shown with the
dark red line in Figure~\ref{fig:limit}. 
Although the limit from GRB090113 is
the best direct observation limit from ANITA-II, it
still suffers from poor geometry.  If the burst had occurred at the angle of 
maximum sensitivity of ANITA-II ($-10^{\circ}$ in elevation), the fluence limit 
would have been $E^4\Phi$=5.2$\times 10 ^{16}$~GeV$^3$cm$^{-2}$.
%the light red line in Figure~\ref{fig:limit}.

There was one downgoing burst (GRB090107A) at an elevation angle of $0.5^{\circ}$ which occurred
while ANITA-II was over the Ross Ice Shelf, allowing for a reflected
observation\footnote{GRB090107A occurred on 2009 January 07 at 04:48:04 UTC, 
with a right ascension of $20^{\mathrm{h}}9^{\mathrm{m}}38^{\mathrm{s}}.16$ and 
a declination of $4^{\circ}44{\mathrm{'}}38~40{\mathrm{''}}$~\citep{gcn}.
The calculated synchrotron break energy is $1.27\times10^{7}$~GeV and the flux above the break
energy is E$^{4}\Phi=9.53\times 10^{6}$~GeV$^3$cm$^{-2}$s$^{-1}$}.  Because of ANITA's proxmity
to McMurdo station during this burst, the background period had significant anthropogenic
noise and the burst was excluded from the sensitive analysis described here.  However,
if we use the same analysis cuts as in the blind diffuse UHE neutrino search described in~\citet{anita2}, we observe 0 events in the prompt and precursor emission window
for this burst.  Although the search sensitivity is worse for this burst because we did 
not loosen the thermal-noise and man-made noise cuts as described above, the observation
of 0 events from this burst still leads to the best limit from ANITA-II on the UHE
neutrino fluence from gamma-ray bursts, $E^4\Phi$=2.5$\times 10 ^{17}$~GeV$^3$cm$^{-2}$, shown 
in blue in Figure~\ref{fig:limit}.
%Figure~\ref{fig:limit} also shows
%the best fluence limits from ANITA-I, RICE, and IceCube, and the Waxman-Bahcall fluence
%calculation for the associated GRBs.

\section{Conclusions}
We have used the temporal information of GRBs which occurred during the
flight of ANITA-II to search for coincident prompt and precursor neutrino emission with greatly
reduced background and improved threshold relative to previous UHE astrophysical
neutrino searches with ANITA.  While the expected fluence based on the standard models of GRB 
particle production is too low to have expected a detection, we present the first limits
on GRB neutrino fluence for energies above $10^9$~GeV.  There is room for about a factor of five
improvement with ANITA-III if a GRB occurs with a good geometry relative to the payload.

\begin{table}[htb2!]
\caption{\label{tab:fluence}Break energies and total flux in the UHE regime
calculated for the 12 bursts in the blind analysis
}
\begin{center}
\begin{tabular}{lccc}
\hline
GRB & $z$ & Synchrotron Break & E$^{4}\Phi$ (GeV$^3$cm${^{-2}}$s$^{-1}$)\\
& & Energy (GeV) & \\
\hline
081228 & 3.8 & 2.18$\times 10^7$ & 2.63$\times 10^8$ \\
081229 & - & 7.93$\times 10^7$ & 8.53$\times 10^{10}$ \\
081230 & - & 1.09$\times 10^7$ & 3.03$\times 10^7$ \\
081231 & - & 5.47$\times 10^7$ & 1.93$\times 10^{10}$ \\
090102 & 1.55 & 1.10$\times 10^6$ & 4.29$\times 10^3$ \\
090108B & - & 7.55$\times 10^7$ & 7.03$\times 10^{10}$ \\
090109 & - & 4.18$\times 10^7$ & 6.60$\times 10^9$\\
090111 & - & 1.22$\times 10^7$ & 4.81$\times 10^7$ \\
090112A & - & 2.40$\times 10^7$ & 7.21$\times 10^8$ \\
090112B & - & 5.59$\times 10^7$ & 2.12$\times 10^{10}$ \\
090113 & - & 1.66$\times 10^7$ & 1.63$\times 10^8$ \\
090117B & - & 2.37$\times 10^7$ & 6.81$\times 10^8$ \\
\hline
\end{tabular}
\end{center}
\end{table}

%\bibliographystyle{elsarticle-num}

\end{document}